\documentclass{article}

\usepackage{epsfig}
\usepackage{graphics}
\usepackage{newsprocl}
\begin{document}

\title{Quantum Chaos Viewed From Quantum Action}

\author{D. Huard, H. Kr\"oger, G. Melkonyan, L.P. Nadeau}
\address{D\'epartement de Physique, Universit\'e Laval, 
Qu\'ebec, Qu\'ebec G1K 7P4, Canada}
\author{K.J.M. Moriarty}
\address{Department of Mathematics, 
Statistics and Computational Science, 
Dalhousie University, Halifax, Nova Scotia B3H 3J5, Canada}

\maketitle

\begin{abstract}
We consider a mixed chaotic Hamiltonian system and compare classical with quantum chaos. As alternative to the methods of energy level spacing statistics 
and trace formulas, we construct a quantum action and a quantum analogue phase space to analyse quantum chaos. 
\end{abstract}

\section{Introduction}
\label{sec:Introduction}

This paper is about the use of the quantum action\cite{kn:Jirari01a,kn:Jirari01b,kn:Caron01,kn:Jirari02,kn:Kroger02,kn:Huard03} in physics. Let us start out by discussing some possible areas of application.

{\bf High energy physics} \\
{\it Renormalisation}. The quantum action may turn out to provide a new definition of renormalisation. In conventional terminology, renormalisation means to extract physical observable parameters like mass, coupling etc. obtained from an interacting quantum field theory at the continuum limit.
Those physical parameters differ from the so-called bare parameters and this difference represents the effects of the interaction. 
Very similar to this, the quantum action has parameters (mass, potential parameters) different from the classical action and this difference 
represents the effects of quantum mechanics: Q.M. fluctuations occuring in the path integral are represented by a single path, however, for a particle with different properties of mass and interaction. 
In order to explore the use of the quantum action as means of renormalisation,
one could construct the quantum action for a many-body system (like a chain of coupled anharmonic oscillators) and explore its quantum mechanical continuum limit. In quantum field theory, one carries out renormalisation by computing n-point vertex functions, which represent vacuum-to-vacuum transition amplitudes for transition time $T \to \infty$.
In the same time limit, the existence of the quantum action has been proven rigorously in Q.M., although so far only in the case of a single particle system. \\   
{\it Cosmology and inflationary scenario of early universe}. 
Inflation involves potentials with several minima and instanton solutions. 
The instanton starts out as a quantum instanton and eventually turns into a classical instanton. This has effects on the subsequent formation of 
galaxies\cite{kn:Staro79,kn:Khlo98,kn:Kolb91}. Using the effective potential or the quantum potential (potential of the quantum action) in general creates a potential different from the classical one. In particular, it may have minima being absent in the classical potential. Consequently, this may create instantons being quite different from the classical instanton. In Ref.\cite{kn:Jirari01b} instantons from the quantum action were found to be "softer" than the corresponding classical instanton. Such quantum effects of the instanton may influence the outcome of the galaxy formation at the end of inflation. \\
{\it Hot and dense nuclear matter}. 
Instantons are believed to play an important role in hot nuclear matter in the quark-gluon plasma phase\cite{kn:Shur88}. 
Instantons are important also for the mechanism of chiral symmetry breaking and for t'Hoofts solution of the U(1) problem. \\
{\it Neutrino oscillations}. 
The process of oscillations of neutrino 
flavors\cite{kn:Pastor02,kn:Horowitz99,kn:Lindner02} may have to do with a process of tunneling in a potential with degenerate minima.

{\bf Condensed matter physics} \\
{\it Quantum dots, semi-conductor and quantum chaos}. 
Advancing the speed of microprocessors may have technological obstacles but also physical limits. When reducing the size of a chip one soon may enter the regime where quantum laws rule. Quantum chaos may become a very important issue, because it can hamper the flow of electric currents. The quantitative determination of quantum chaos effects will possibly be of great importance for the development of future microprocessors. 
By use of some kind of effective action or the quantum action, one can study  quantum corrals formed by atoms and quantum dots in semi-conductors.  
In particular, this allows to study the temperature dependence of electron dynamics in atomic corrals, as well as for electrons moving in simple conductor-semiconductor-isolator geometries and to search for the possible presence of quantum chaos. \\
{\it Josephson junctions and superconducting quantum interference}. Superconducting quantum interference devices (SQUID)
have been used to demonstrate experimentally the phenomenon of quantum superposition in macroscopic states\cite{kn:Friedman00}. This involves Josephson junctions. The SQUID potential has a double-well structure. 
The effective potential and the quantum potential should be useful tools to analyze quantum superposition in terms of such potential involving quantum effects. \\ 
{\it Quantum computers based on superconductors}.
The symmetry of the order parameter in some triplet superconductors corrersponds to doubly degenerate chiral states.
Gulian et al.\cite{kn:Gulian02a,kn:Gulian02b} predict that this degeneracy can be lifted via macroscopic quantum tunneling. Instanton-like quantum behavior may become important. Triplet superconducters may be used as basic elements of quantum computers. Again the effective potential and the quantum potential should help to study quantum instantons and tunneling in such materials.

{\bf Atomic physics} \\
{\it Analogy of classical chaos in quantum physics}. 
Attempts have been made to use the effective action in order to characterize chaotic behavior in quantum systems\cite{kn:JonaLas00}. In a a similar way, the quantum action has been used also\cite{kn:Caron01}. This allows to construct of a phase space portrait and Poincar\'e sections for a quantum system in analogy to classical physics. From this one can obtain Lyapunov exponents and KAM surfaces for the quantum system. A potentially quite interesting system to explore is the Paul trap. \\
{\it Ultracold atoms in a billard formed by lasers}. 
Trace formulas (Gutzwiller\cite{kn:Gutzwiller90} and generalisations ) have been used successfully\cite{kn:Milner01,kn:Friedman01,kn:Dembrowski01} to establish a relation between level densities and periodic semi-classical orbits. It would be interesting to compare predictions of trace formulas in the semi-classical regime with the predictions obtained from the effective action or the quantum action. \\
{\it Dynamical tunneling}. 
Steck et al.\cite{kn:Raizen01} and Hensinger et al.\cite{kn:Hensinger01}  
have demonstrated experimentally the phenomenon of dynamical tunneling (where the classical transition is forbidden due to some conserved quantity different from energy). It has been realized by arrays of cold atoms. It has been observed that the presence of quantum chaos enhances the dynamical tunneling transition.
It would be instructive to reexamine dynamical tunneling using the phase space portrait constructed from a time-dependent effective action or quantum action.

{\bf Chemistry} \\
{\it Binding of macromolecules}. 
In the process of chemical binding of macromolecules, often a double well potential plays a role. The effective action as well as the quantum action should be useful to find pathways in the formation of such macromolecules.

This paper is focussed on the quantum 
action\cite{kn:Jirari01a,kn:Jirari01b,kn:Caron01,kn:Jirari02,kn:Kroger02,kn:Huard03}. We discuss some of its analytical properties. We present results of numerical computations testing the goodness of the quantum action in fitting transition amplitudes. As a physical application we address the topic of quantum chaos, occuring in nuclear physics and atomic physics, from the viewpoint of the quantum action. The quantum action parametrizes quantum transition amplitudes $G$ in terms of a new action $\tilde{S}$ defined by 
\begin{eqnarray}
\label{eq:DefQuantAct}
&& G(x_{fi},t=T;x_{in},t=0) = 
\sum_{x_{traj}} \tilde{Z}_{x_{traj}} \exp[ i \tilde{S}[x_{traj}]/\hbar ] 
\nonumber \\
&& \tilde{S}[x] =
\int_{0}^{T} dt \frac{1}{2} \tilde{m} \dot{x}^{2} - \tilde{V}[x] ~ ,
\end{eqnarray}
where $\tilde{S}$ is a local action - the so called quantum action and $x_{traj}$ is the classical trajectory of such action going from boundary point $(x_{in},t=0)$ to $(x_{fi},t=T)$. For some potentials there may be multiple trajectories connecting those boundary points, then there is a sum over all trajectories. $\tilde{Z}$ is a normalisation factor. One should note that the existence of such quantum action in general has not been proven yet. As an operational definition of the quantum action, one may consider it as that action which gives the optimal fit of the transition amplitudes $G$ for all combinations of boundary points, for a fixed transition time. Often such optimal fit using only a single trajectory is already very good, as can be seen in examples below.

In the limit of large (imaginary) transition time the quantum action has been proven rigorously to exist, giving an exact parametrisation of transition amplitudes\cite{kn:Kroger02}. The quantum action being local allows to construct a portrait of phase space of a quantum system by applying the tools of nonlinear dynamics to this action\cite{kn:Caron01}. In particular, it allows to construct Poincar\'e sections and Lyapunov exponents, in analogy to classical nonlinear dynamics (note that the definition of Lyapunov exponents requires the limit $T \to \infty$). 
\begin{figure}[thb]
\vspace{9pt}
\begin{center}
\includegraphics[scale=0.35,angle=0]{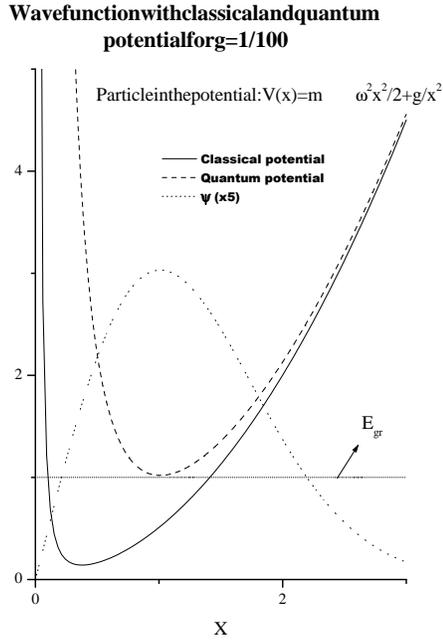}
\end{center}
\caption{Inverse square potential. Classical potential with parameters $m=1$, $\omega=1$ and $g=10^{-2}$ (full line), quantum potential (dashed line) and ground state wave function (enhanced by factor 5, dotted line). }
\label{fig:PotWave}
\end{figure}

\section{Analytical properties of quantum action}
\label{sec:Analytic}
Consider the Schr\"odinger equation in imaginary time ($ t \to -it$) with a given time-independent Hamiltonian $H$. Then the transition amplitude evolves according to
\begin{eqnarray}
&& G(x_{fi},T;x_{in},0) \sim_{T \to \infty} = 
\sum_{n} <x_{fi}|\psi_{n}> \exp[-E_{gr}T/\hbar] <\psi_{n}|x_{in}> 
\nonumber \\
&=& \sim_{T \to \infty} ~ <x_{fi}|\psi_{gr}> \exp[-E_{gr}T/\hbar] <\psi_{gr}|x_{in}> ~ .
\end{eqnarray}
In the large time limit (Feynman-Kac limit) the transition amplitude  
is projected onto the ground state contribution. This
may give the impression that for large time most of the physical 
information is lost. However, this is not true. The stationary Schr\"odinger equation allows to reconstruct the potential $\times$ mass from the ground state
\begin{equation}
\label{eq:PsiClassPot}
\frac{\Delta \psi_{gr}}{\psi_{gr}} = \frac{2m}{\hbar^{2}} (V-E_{gr}) ~ .
\end{equation}
The mass can be reconstructed from the groundstate plus any one (e.g. lowest) of the excited states,
\begin{equation}
\frac{\Delta \psi_{ex}}{\psi_{ex}} - 
\frac{\Delta \psi_{gr}}{\psi_{gr}} =
- \frac{2m}{\hbar^{2}} (E_{ex}-E_{gr}) ~ .
\end{equation}
In other words, all information on the Hamiltonian is stored in the wave functions and energies of any two eigen states. This is to some extent 
analogous to biology, where the genetic code is stored in all cells. 
Similar to Eq.(\ref{eq:PsiClassPot}) there is a relation between the ground state and the quantum potential, proven in Ref.\cite{kn:Kroger02},
\begin{equation}
\label{eq:PsiQuantPot}
\left( \frac{\vec{\nabla} \psi_{gr}}{\psi_{gr}} \right)^{2} = 
\frac{2 \tilde{m}}{\hbar^{2}} (\tilde{V} - \tilde{v}_{min}) ~ , ~~~ \tilde{v}_{min} = E_{gr} ~ ,
\end{equation}
where $\tilde{v}_{min}$ denotes the minimum value of the quantum potential. In $D=1$ dimension, this allows to express the wave function in terms of the quantum potential and quantum mass 
\begin{equation}
\label{GroundStateLaw}
\psi_{gr}(x) = \frac{1}{N} ~ \exp[ - \int_{0}^{|x|} dx' ~ 
\sqrt{2 \tilde{m}( \tilde{V}(x') - \tilde{v}_{min} ) }/\hbar ] ~ .
\end{equation}
This expression holds in the case when the classical potential is parity symmetric, has a single minimum at $x=0$ and the ground state wave function a single maximum at $x=0$. 
The formula can be generalized to multi-valley potentials. 
Now we will establish a differential equation relating the classical potential to the quantum potential.
Defining
\begin{equation}
\label{eq:DefU}
\vec{U} = \vec{\nabla} \log \psi_{gr} ~ ,
\end{equation}
the combination of Eqs.(\ref{eq:PsiClassPot},\ref{eq:PsiQuantPot},\ref{eq:DefU}) yields
\begin{equation}
\label{eq:Riccati}
\vec{U}^{2} + \vec{\nabla} \cdot \vec{U} = 
\frac{2 m}{\hbar^{2}} (V - E_{gr}) ~ , ~~~~
\vec{U}^{2} = \frac{2 \tilde{m}}{\hbar^{2}} (\tilde{V} - \tilde{v}_{min}) ~ .
\end{equation}
This is a Riccati-type differential equation to determine the function $\frac{2 \tilde{m}}{\hbar^{2}} (\tilde{V} - \tilde{v}_{min})$ when 
$\frac{2 m}{\hbar^{2}} (V - E_{gr})$ is given. In $D=1$ it can be expressed more explicitely by
\begin{equation}  
\label{eq:TransformLaw} 
2 \tilde{m}(\tilde{V}(x) - \tilde{v}_{0}) 
- \frac{\hbar}{2} \frac{ \frac{d}{dx} 2 \tilde{m} (\tilde{V}(x) - \tilde{v}_{0})}
{ \sqrt{2 \tilde{m}( \tilde{V}(x) - \tilde{v}_{0} ) } } ~ \mbox{sgn}(x) 
= 2 m(V(x) - E_{gr}) ~ .
\end{equation}
\begin{figure}[thb]
\vspace{9pt}
\begin{center}
\includegraphics[scale=0.40,angle=270]{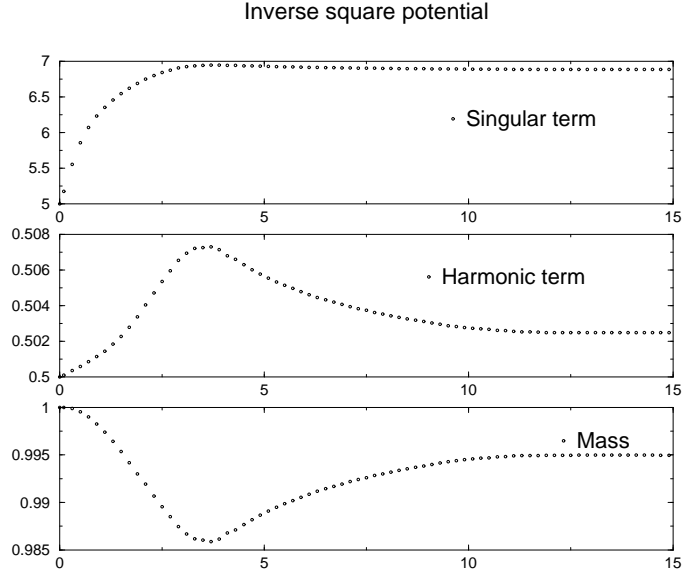}
\end{center}
\caption{Inverse square potential. Parameters of the quantum action 
$\tilde{m}$ (bottom), $\tilde{v}_{2}$ (center) and $\tilde{v}_{-2}$ (top) as function of transition time $T$. 
Classical parameters $m=1$, $v_{2}=0.5$ and $v_{-2}=5$. }
\label{fig:ActPar}
\end{figure}

\section{Quantum action versus supersymmetric partner potential}
\label{sec:SuperSymm}
In $D=1$ ideas from supersymmetry have been used to improve the computation of tunneling amplitudes\cite{kn:Keung88,kn:Kumar86}. For a given potential, there is a supersymmetric partner potential, which yields the same spectrum as the original potential, except for the ground state.
Consider a Hamiltonian of the form (taking $2m=\hbar=1$),
\begin{equation}
H = - \frac{d^{2}}{dx^{2}} + V_{-}(x) ~ , ~~~ V_{-} \equiv V ~ ,
\end{equation}
there is a supersymmetric partner potential $V_{+}$, given by
\begin{equation}
V_{+}(x) = - V_{-}(x) + 2 \left( \frac{\frac{d}{dx} \psi_{gr}(x)}{\psi_{gr}(x)} \right)^{2} ~ .
\end{equation}
Defining 
\begin{equation}
W(x) = \frac{\frac{d}{dx} \psi_{gr}(x)}{\psi_{gr}(x)} ~ , ~~~ 
W^{2} = \frac{1}{2} (V_{+}+V_{-}) ~ ,
\end{equation}
yields the relation
\begin{equation}
\label{eq:W2}
W^{2} = \tilde{V} - \tilde{v}_{min} ~ , 
\end{equation}
and obeys the Riccati differential equation
\begin{equation}
\label{eq:Riccati1D}
W^{2}(x) + \frac{dW}{dx} = V(x) - E_{gr} ~ .
\end{equation}
Eqs.(\ref{eq:W2},\ref{eq:Riccati1D}) are identical to Eq.(\ref{eq:Riccati}) in $D=1$. As a lesson, one finds that the formalism of the supersymmetric partner potential has a close correspondence in $D=1$ to the quantum action. 
The concept of the supersymmetric partner potential and the quantum action, however, differ in two respects: First The supersymmetric partner potential has no physical interpretation in contrast to the quantum action, which gives a representation of Q.M. transition amplitudes. Second, the concept of the supersymmetric partner potential is restricted to $D=1$ 
dimension\cite{kn:Kumar86}, in contrast to the quantum action which is valid also in $D=2,3$ dimensions. It would be tempting to apply the quantum action to the tunneling problem in $D=2,3$ dimensions. Some work is in progress.
\begin{figure}[thb]
\vspace{9pt}
\begin{center}
\includegraphics[scale=0.40,angle=270]{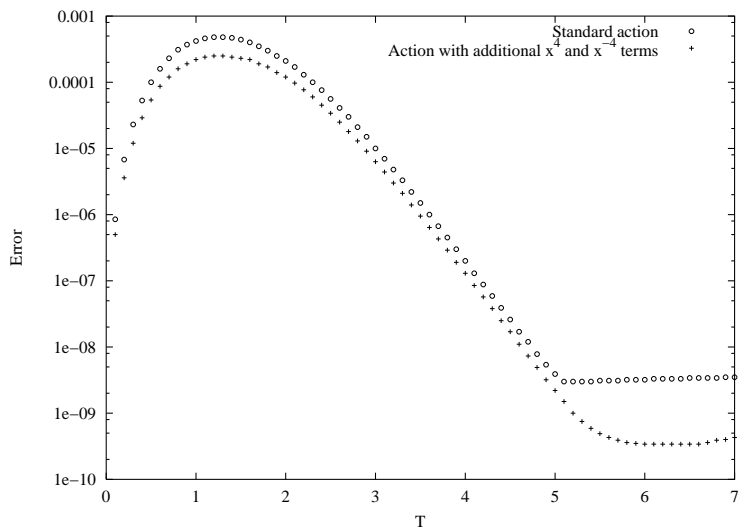}
\end{center}
\caption{Global relative error of quantum action $\Sigma_{ij}$. }
\label{fig:e44_S}
\end{figure}

\section{Test of quantum action: inverse square potential}
\label{Sec:Test}
\begin{figure}[thb]
\vspace{9pt}
\begin{center}
\includegraphics[scale=0.40,angle=270]{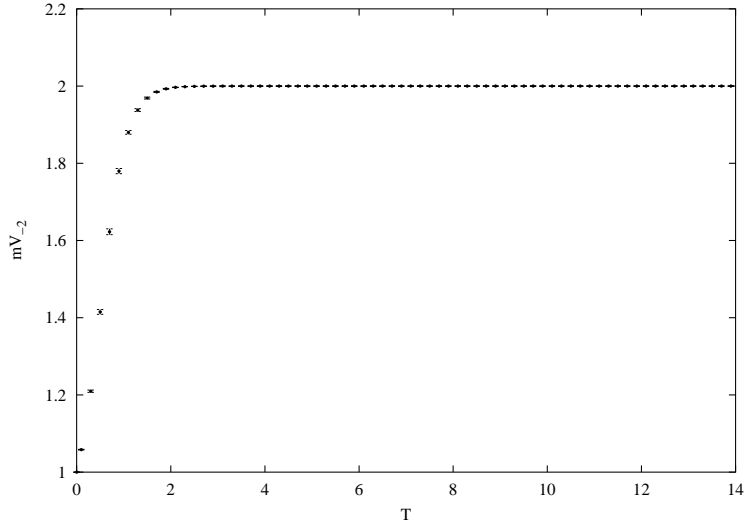}
\end{center}
\caption{Quantum action parameters $\tilde{m} \tilde{v}_{-2}$ obtained from global fit method vs. transition time $T$.}
\label{fig:estand_mv-2}
\end{figure}

\begin{figure}[thb]
\vspace{9pt}
\begin{center}
\includegraphics[scale=0.40,angle=270]{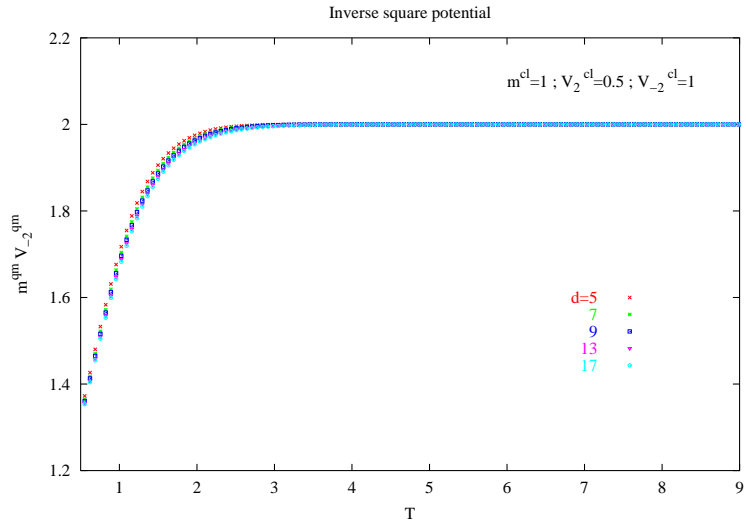}
\end{center}
\caption{Quantum action parameter $\tilde{m} \tilde{v}_{-2}$ obtained from flow equation vs. transition time $T$. Dependence on location of interval $[c,d]$ of final boundary points. }
\label{fig:eqdiff_mv-2}
\end{figure}

We consider the following classical potential in 1-D
\begin{equation}
\label{ClassPot}
V(x) = \frac{1}{2} m \omega^{2} ~ x^{2} + g ~ x^{-2} 
= v_{2} ~ x^{2} + v_{-2} ~ x^{-2} ~ .
\end{equation}
The corresponding classical action is given by
$S = \int dt ~ \frac{1}{2} m \dot{x}^{2} - V(x)$.
The potential is parity symmetric.
Because it has an infinite barrier (for $g > 0$) at the origin,
the system at $x < 0$ is separated from the system at $x > 0$. We consider only the motion in the domain $x > 0$. We have chosen to consider the inverse square potential, because of the distinct feature that the corresponding quantum mechanical transition amplitudes are known 
analytically\cite{kn:Khandekar75,kn:Schulman81}. 
The classical potential compared to the quantum potential as well as the ground state wave function is shown in Fig.[\ref{fig:PotWave}]. 
The behavior of the parameters of the quantum action as a function of transition time $T$ are shown in Fig.[\ref{fig:ActPar}]. 
The goodness of the fit of transition amplitudes by the quantum
action, using only one trajectory is shown in Fig.[\ref{fig:e44_S}].
It shows the global relative error of the quantum action versus transition time. One observes a maximal error in the order of $4 \times 10^{-4}$
for $T \approx 1$. When $T$ approaches zero, the error goes to zero. It becomes quite small also for $T \to \infty$, consistent with the analytic behavior of the quantum action. Moreover, the figures shows that the error decreases, if the quantum potential allows for the presence of terms like $x^{4}$ and $x^{-4}$, which are absent in the classical potential. 
Fig.[\ref{fig:estand_mv-2}] shows the behavior of the parameter combination $\tilde{m} \tilde{v}_{-2}$ versus transition time, computed by doing an optimisation of the quantum action parameters to globally fit the transition amplitudes.
Fig.[\ref{fig:eqdiff_mv-2}] shows the same quantity, computed in an alternative way from the flow equation, discussed in Ref.\cite{kn:Huard03}. One observes that both methods yield asymptotically stable results which agree among each other and their asymptotic values agree also with the analytical value $\tilde{m} \tilde{v}_{-2} \to_{T \to \infty} 2$, predicted by Eq.(\ref{eq:TransformLaw}).

\section{Quantum chaos}
\label{sec:Chaos}
Quantum chaos has been experimentally observed in irregular 
energy spectra of nuclei, and also in atoms exposed to strong electromagnetic fields\cite{kn:Friedrich89}. Classically chaotic billiard systems have a quantum counterpart being atoms moving in a 2-D billiard, where bounday reflections are caused by controlled interaction with laser 
beams\cite{kn:Milner01,kn:Friedman01,kn:Dembrowski01}. Irregular patterns have been predicted by computer simulations and were found experimentally in wave functions of quantum billards\cite{kn:McDonald79}, where scars are reminders of classical motion\cite{kn:Heller84}.
\begin{figure}[tph]
\vspace{9pt}
\begin{center}
\includegraphics[scale=0.30,angle=0]{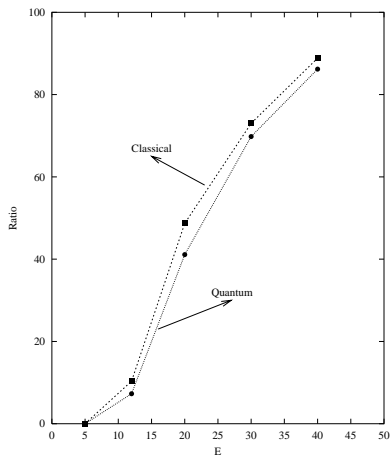}
\end{center}
\caption{Volume of chaotic phase space (positive Lyapunov exponent) over total phase space (lines are guides to the eye). $v_{22}=0.05$.}
\label{fig:Ratio}
\end{figure}
Recently, dynamical tunneling in atoms between regular islands has been observed. The transition is enhanced by chaos\cite{kn:Raizen01,kn:Hensinger01}.
Spectra of fully chaotic quantum systems can statistically be described by random matrices\cite{kn:Bohigas84}, which corresponds to a level density distribution of Wigner-type, while integrable (nonchaotic) quantum system 
yield a Poissonian distribution. Here we ask: What happens between these two extremes? For example, an hydrogen atom exposed to a weak exterior magnetic field shows a level distribution, which is neither Poissonian nor Wignerian. Can we compare classical chaos with quantum chaos?
Is the quantum system more or less chaotic than the corresponding classical system? Moreover, an understanding of how classically mixed phase space is reflected in quantum systems is an open problem. Semiclassical methods of quantisation (EKB, Gutzwiller's trace formula) are not amenable to mixed dynamical systems\cite{kn:Bohigas93}.
Here we want to address those questions using the concept of the quantum 
action\cite{kn:Jirari01a,kn:Jirari01b,kn:Caron01,kn:Jirari02,kn:Kroger02,kn:Huard03}.
We have considered the following $2-D$ Hamilton system, which is classically chaotic,
\begin{equation} 
S = \int_{0}^{T} dt ~ \frac{1}{2} m (\dot{x}^{2} + \dot{y}^{2}) - V(x,y) ~ , ~~~ V = v_{2}(x^{2} + y^{2}) + v_{22} x^{2}y^{2} ~ .
\end{equation}
In the quantum action we made an ansatz of the following form for the quantum potential,
\begin{eqnarray}
&& \tilde{S} = \int_{0}^{T} dt ~ \frac{1}{2} \tilde{m} (\dot{x}^{2} + \dot{y}^{2}) - \left\{ 
 \tilde{V}(x,y) = \tilde{v}_{0} + \tilde{v}_{11} xy + \tilde{v}_{2} (x^{2}+y^{2}) + \tilde{v}_{22} x^{2}y^{2} \right.
\nonumber \\
&& \left. + \tilde{v}_{13} (xy^{3} + x^{3}y)
+ \tilde{v}_{4} (x^{4} + y^{4}) + \tilde{v}_{24} (x^{2}y^{4} + x^{4}y^{2})
+ \tilde{v}_{44} x^{4}y^{4} \right\} ~ .
\end{eqnarray}
The parameters of the quantum action have been determined by a global fit to transition amplitudes. 
The terms $\tilde{v}_{11}$, $\tilde{v}_{13}$, $\tilde{v}_{4}$, $\tilde{v}_{24}$,  $\tilde{v}_{44}$, were found to be quite small or compatible with zero.
A comparison of Poincar'e sections 
of the classical action and the quantum action shows that there are slightly more and larger regular islands in the quantum case than in the classical case. The ratio of chaotic phase space is displayed in Fig.[\ref{fig:Ratio}]. It shows that overall the quantum system is less chaotic. \\

H.K. and K.J.M.M. are grateful for support by NSERC Canada. \\

\end{document}